\documentclass[sigplan,10pt]{acmart}

\renewcommand\footnotetextcopyrightpermission[1]{}
\usepackage{graphicx,subcaption,pifont}
\usepackage{algorithmic}
\usepackage{algorithm}
\usepackage{multirow}
\usepackage{ulem}
\usepackage{booktabs}
\usepackage{array}
\usepackage{diagbox}
\usepackage{tabularx}
\usepackage{paralist}
\usepackage{tikz}
\usepackage{amsmath}
\usepackage{comment}
\usepackage{xurl}
\usepackage{filecontents}

\def\myvar{Mole}

\begin{document}

\title{Mole: Tier-Specific Hotness Profiling Driven Memory Tiering for Multi-Tiered Memory Systems}

\author{Mingyang Liu}
\affiliation{
  \institution{Xiamen University}
  \city{Xiamen}
  \country{China}
}
\email{mingyang@stu.xmu.edu.cn}

\author{Congming Gao}
\affiliation{
  \institution{Xiamen University}
  \city{Xiamen}
  \country{China}
}
\email{gaocm92@gmail.com}

\author{Xufeng Yang}
\affiliation{
  \institution{Xiamen University}
  \city{Xiamen}
  \country{China}
}
\email{xufengyang@stu.xmu.edu.cn}

\author{Fang Wu}
\affiliation{
  \institution{Xiamen University}
  \city{Xiamen}
  \country{China}
}
\email{wfiu666666@gmail.com}

\author{Youmin Chen}
\affiliation{
  \institution{Shanghai Jiao Tong University}
  \city{Shanghai}
  \country{China}
}
\email{chenyoumin@sjtu.edu.cn}

\author{Renhui Chen}
\affiliation{
  \institution{Xiamen University}
  \city{Xiamen}
  \country{China}
}
\email{chenrenhui@stu.xmu.edu.cn}

\author{Jiwu Shu}
\affiliation{
  \institution{Xiamen University}
  \city{Xiamen}
  \country{China}
}
\email{jwshu@xmu.edu.cn}

\begin{abstract}
Multi-tiered memory systems combine fast, small upper tiers with slow, large lower tiers to improve performance and cost efficiency. Existing designs, such as AutoTiering and MTM, rely on greedy promotion and stepwise demotion, which can intensify contention for limited capacity in faster tiers. We observe that promotions directly improve performance, whereas demotions primarily reclaim space. Based on this asymmetry, we propose Mole, a memory tiering system that separates promotion and demotion destinations. Mole employs staging demotion to move cold pages directly to the lowest tier, bypassing intermediate tiers, and targeted promotion to place hot pages in tiers that match their current hotness. The lowest tier serves as a staging area from which pages can be promoted when they become hot again. This separation preserves intermediate-tier capacity for promotions, reducing tier contention and migration failures. However, staging demotion requires timely identification of pages that become hot after demotion. To meet this requirement with low profiling overhead, Mole uses tier-specific profiling: it profiles the lowest tier at high frequency to detect reactivated hot pages, while profiling upper tiers at lower frequency to identify cold pages. Experimental results show that Mole reduces migration failures and improves performance under dynamic access patterns.
\end{abstract}

\settopmatter{printfolios=true}
\pagestyle{plain}

\maketitle

\section{Introduction}\label{sec:intro}
Existing memory systems commonly adopt a tiered hierarchy from top to bottom: the top tier is fast, expensive, and small, used for frequently accessed (hot) pages; the bottom tier is slow, cheap, and large, used for cold pages. With advanced memory technologies like persistent memory (PM)~\cite{PM_2019} and CXL-attached memory~\cite{cxl_members, CXL_Test_MICRO_23, NeoMem_MICRO_24, Spa_asplos_26}, memory hierarchies are scaling out to more tiers, enabling new tiered, heterogeneous memory systems~\cite{Pond_ASPLOS_23, TPP_ASPLOS23, MEMTIS_SOSP_23, Chrono_EuroSys2025, PET_EuroSys25}. In multi-tiered memory systems, pages must be placed on the tier that best matches their access hotness. This triggers data migration across tiers, generating substantial tier-transit traffic and new challenges.

\begin{figure}[t]
    \centering
    \begin{subfigure}[b]{0.23\textwidth}
        \includegraphics[width=\textwidth]{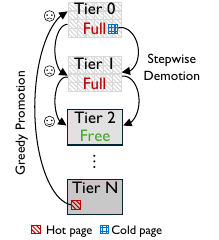}
        \caption{Greedy promotion in \cite{AutoTiering_ATC_21, MTM_EuroSys_24}.}
        \label{fig:intro:greedy_promotion}
    \end{subfigure}
    \hfill
    \begin{subfigure}[b]{0.23\textwidth}
        \includegraphics[width=\textwidth]{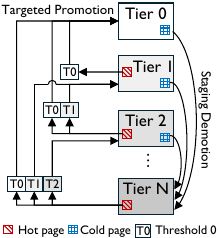}
        \caption{Staging demotion in Mole.}
        \label{fig:intro:aggressive_demotion}
    \end{subfigure}
    \vspace{-0.1in}
    \caption{Migration policy in existing multi-tier memory tiering systems versus Mole.}
    \label{fig:intro:greedy_promotion_and_staging_demotion}
    \vspace{-0.25in}
\end{figure}

Existing multi-tiered memory management systems, including AutoTiering~\cite{AutoTiering_ATC_21} and MTM~\cite{MTM_EuroSys_24}, employ a \textbf{\textit{greedy promotion}} policy for moving pages across tiers.
Greedy promotion always attempts to place hot pages into the top tier, falling back to the next available tier if full, while \textbf{\textit{stepwise demotion}} moves cold pages downward until finding available capacity.
As illustrated in Figure~\ref{fig:intro:greedy_promotion}, greedy promotion prioritizes placing hot pages in the top tier (Tier-0). 
If Tier-0 lacks sufficient capacity, it continues searching downward through the hierarchy until it finds a tier with available capacity (i.e., Tier-2).
Stepwise demotion is triggered when a tier’s occupied pages exceed a predefined watermark, after which identified cold pages in that tier (i.e., Tier-0) are moved down the hierarchy, skipping fully occupied tiers, until reaching the first tier with available capacity (i.e., Tier-2).
We refer to competition among pages for limited capacity within a memory tier as \textbf{tier contention}.
Greedy promotion and stepwise demotion can intensify this contention by increasing demand for limited capacity in faster tiers.
When destination tiers lack sufficient available capacity, promotions and demotions may fail, leaving pages in suboptimal tiers.
Experiments in \S\ref{sec:motiv:migration_fail} reveal promotion failure rates above 97\% for both Silo and Liblinear under saturated conditions, and demotion failure rates up to 37.9\% and 48.2\%, respectively.
These failures occur even when many hot pages are correctly identified, indicating that accurate hotness estimation alone is insufficient to ensure successful migration under capacity pressure.

When tier capacity is limited, promotions and demotions should not be treated symmetrically.
Promotions move frequently accessed pages toward faster tiers and directly improve performance, whereas demotions mainly reclaim space from pages with lower expected reuse.
Therefore, we let demotions yield to promotions by bypassing intermediate tiers and moving cold pages directly to the lowest tier.
Based on this, we design \textbf{Mole} with \textbf{staging demotion} and \textbf{targeted promotion}.
As shown in Figure~\ref{fig:intro:aggressive_demotion}, Mole separates the destinations of promotions and demotions. 
Promotions use dynamic hotness thresholds to target the tier that matches each page's current hotness, including the top and intermediate tiers (Tiers-0$\sim$N-1). 
Demotions, in contrast, bypass intermediate tiers and move cold pages directly to the lowest tier (Tier-N), which serves as a staging area for pages that may later warrant promotion.
By redirecting migration destinations, Mole prevents demotions from occupying promotion target tiers, preserving capacity for future promotions and ensuring migration success rate.

However, staging demotion is not free.
Page hotness is dynamic, and demoted cold pages may later become hot again.
Delayed hot page detection in the lowest tier can prevent these reactivated pages from being promoted in time, thus increasing the access latency of hot pages and amplifying the penalty of staging demotion.
Therefore, staging demotion requires an accurate and timely hot page hotness detection in the lowest tier.
A natural way to improve detection timeliness is to increase page hotness profiling frequency.
However, high-frequency profiling incurs substantial overhead~\cite{MEMTIS_SOSP_23, Telescope_ATC_24} and cannot be applied uniformly across all tiers, especially in the upper tier, where most pages are hot and their performance can be significantly affected by frequent profiling. 
Fortunately, upper tiers already contain most hot pages, so there is no need to repeatedly detect their hotness; besides, accurately identifying cold pages in the upper tier does not require high-frequency profiling as well: cold pages have long access intervals, so a moderate detection delay in identifying them has limited impact on overall hotness identification. 
Consequently, staging demotion can be harnessed with an asymmetrical profiling frequency across tiers: \textit{high-frequency profiling on lower tiers promptly detects reactivated pages with minimal impact on cold pages, while low-frequency profiling on the upper tier identifies cold pages without significantly affecting hot pages}.

However, existing profiling mechanisms do not fully exploit this asymmetry.
Most existing systems use an identical profiling frequency across tiers~\cite{AutoTiering_ATC_21, MTM_EuroSys_24, Telescope_ATC_24}, which couples the detection latency of the lowest tier with the profiling overhead incurred in upper tiers. 
Although several prior works~\cite{TPP_ASPLOS23, NeoMem_MICRO_24, AutoNUMA_lwn} have exploited this asymmetry by applying profiling only to the lower or remote tier, they lack scalability and cannot support differentiated profiling frequency control on a per‑tier basis that accounts for variations in hotness and page sensitivity across tiers.

To address this limitation, we design \textbf{\myvar{}}, a memory tiering system that combines tier-specific profiling, staging demotion, and targeted promotion.
\myvar{} uses tier-specific profiling to promptly detect misplaced or reactivated pages on lowest tier, and identify cold pages on upper tiers with low overhead.
Based on the collected hotness information, \myvar{} enables staging demotion to move cold pages directly to the lowest tier and targeted promotion to place hot pages into hotness-matched upper or intermediate tiers.
This destination separation prevents demoted cold pages from occupying tiers needed by future promotions, thereby reducing tier contention and improving migration success.
To support tier-specific profiling efficiently, \myvar{} extends hint-page-fault-based profiling with physical-address-based scans over Linux's tier-specific LRU lists.
This allows \myvar{} to select pages from each tier directly, while a page metadata flag records whether a page has already been protected for hint-page-fault profiling, avoiding redundant page table walks.
Besides, \myvar{} measures both unaccessed intervals and reaccess intervals as unified hotness metric across tiers to improve hotness detection accuracy.

The contributions of this work include:
1) Identify the limitations of existing memory tiering in multi-tiered systems: page migration frequently fails, and hotness profiling struggles to simultaneously achieve high accuracy and low overhead;
2) Design \myvar{} to address these limitations: it improves migration success by reducing tier contention and improves profiling efficiency by adapting hotness detection to different tiers;
3) Evaluate \myvar~ on a four-tier memory system. Experimental results show that \myvar~ outperforms state-of-the-art multi-tiered memory tiering systems.

\section{Background and Related Works}\label{Background}

\subsection{Multi-Tiered Memory System}
Conventional two-tiered memory systems combine DRAM with emerging memory technologies like PM\cite{PM_2019}, HBM, or CXL devices\cite{cxl_members}. 
Memory tiers are often deployed with multi-socket NUMA architectures, and CXL switches commonly connect to multiple devices~\cite{NeoMem_MICRO_24, Tigon_OSDI_25}, making multi-tiered systems increasingly prevalent. 
These systems offer scalable, cost-effective capacity for data centers and AI workloads, but also pose challenges due to heterogeneity and cross-socket latency. 
This work focuses on a four-tier memory hierarchy with heterogeneous latencies and capacities. 
We instantiate it on a dual-socket machine, where the memory attached to each socket is exposed as separate NUMA nodes and configured as four tiers~\cite{AutoTiering_ATC_21, MTM_EuroSys_24}. 
To isolate page placement from thread scheduling and cross-socket effects, we bind application threads to a single socket throughout. 
For clarity, since memory tiers are often represented as NUMA nodes~\cite{Memstrata_OSDI24, TPP_ASPLOS23, Pond_ASPLOS_23}, this paper primarily uses the term ``tier'' to focus discussion on tiered memory systems.

\begin{table*}
    \centering
    \caption{Comparison of profiling and migration policies in tiered memory systems.}
    \vspace{-0.15in}
    \newcolumntype{M}[1]{>{\centering\arraybackslash}m{#1}}
    \footnotesize
    \begin{tabular}{|p{1.8cm}M{2cm}|M{3.8cm}M{3.7cm}|M{2.7cm}M{2.8cm}|}
        \hline
        \multicolumn{2}{|c|}{\multirow{2}{*}{}} & 
        \multicolumn{2}{c|}{\textbf{Page Hotness Profiling}} & 
        \multicolumn{2}{c|}{\textbf{Migration Path Management}} \\ 
        \cline{3-6}
        
        \multicolumn{2}{|c|}{} & 
        Method & 
        Profiling Control & 
        Promotion & 
        Demotion \\ 
        \hline
        
        \multicolumn{1}{|c|}{\multirow{10}{*}{\rotatebox[origin=c]{90}{\textbf{Two-Tier}}}} &
        HeMem~\cite{HeMem_SOSP21} &
        PEBS &
        Identical Rate &
        - &
        - \\
        
        \multicolumn{1}{|c|}{} &
        TPP~\cite{TPP_ASPLOS23} &
        PEBS + Hint Page Fault &
        Only Lower Tier &
        - &
        - \\
        
        \multicolumn{1}{|c|}{} & 
        Telescope~\cite{Telescope_ATC_24} & 
        PTE Access Bit & 
        Identical Rate & 
        - & 
        - \\
        
        \multicolumn{1}{|c|}{} & 
        TMTS~\cite{TMTS_ASPLOS_23} & 
        PEBS + PTE Access Bit & 
        Identical Rate & 
        - & 
        - \\
        
        \multicolumn{1}{|c|}{} & 
        NeoMem~\cite{NeoMem_MICRO_24} & 
        Hardware Profiler & 
        CXL Tier & 
        - & 
        - \\
        
        \multicolumn{1}{|c|}{} & 
        MEMTIS~\cite{MEMTIS_SOSP_23} & 
        PEBS & 
        Identical Rate & 
        - & 
        - \\
        
        \multicolumn{1}{|c|}{} & 
        FlexMem~\cite{FlexMem_ATC_24} & 
        PEBS + Hint Page Fault & 
        Identical Rate & 
        - & 
        - \\ 

        \multicolumn{1}{|c|}{} & 
        PET~\cite{PET_EuroSys25} & 
        PTE Access Bit + Hint Page Fault & 
        Identical Rate & 
        - & 
        - \\ 
        \multicolumn{1}{|c|}{} & 
        Chrono~\cite{Chrono_EuroSys2025} & 
        Hint Page Fault & 
        Identical Rate & 
        - & 
        - \\ 

        \multicolumn{1}{|c|}{} & 
        HybridTier~\cite{HybridTier_asplos26} & 
        PEBS & 
        Identical Rate & 
        - & 
        - \\ 
        
        \hline
        
        \multicolumn{1}{|c|}{\multirow{5}{*}{\rotatebox[origin=c]{90}{\textbf{Multi-Tier}}}} & 
        AutoNUMA~\cite{AutoNUMA_lwn} & 
        Hint Page Fault & 
        Only Remote Tier & 
        Only Top Tier & 
        Local Lower Tier \\
        
        \multicolumn{1}{|c|}{} & 
        Tiering-0.8~\cite{Tiering_git} & 
        Hint Page Fault & 
        Identical Rate & 
        Only Top Tier & 
        Local Lower Tier \\
        
        \multicolumn{1}{|c|}{} & 
        AutoTiering~\cite{AutoTiering_ATC_21} & 
        Hint Page Fault & 
        Identical Rate & 
        Best Effort to Top Tier & 
        Lower-performance Tier \\
        
        \multicolumn{1}{|c|}{} & 
        MTM~\cite{MTM_EuroSys_24} & 
        PTE Access Bit & 
        Identical Rate & 
        Best Effort to Top Tier & 
        Lower-performance Tier \\ 
        \cline{2-6}
        
        \multicolumn{1}{|c|}{} & 
        \textbf{\myvar} & 
        \textbf{Hint Page Fault + Scanning} & 
        \textbf{Tier-Specific} & 
        \textbf{Targeted Tier} & 
        \textbf{Lowest Tier} \\ 
        \hline
    \end{tabular}
    \text{\small\textit{Note: \textbf{Identical Rate} means all tiers use the same profiling frequency; ``-'' denotes mechanisms designed for two-tiered memory systems.}}
    \label{tab:summary_of_literature}
     \vspace{-0.15in}
\end{table*}

\subsection{Tiered Memory Management}
Memory tiering comprises two key components: page hotness profiling and migration path management. 
Page hotness profiling is responsible for detecting page access patterns, thereby guiding subsequent migration operations.

\subsubsection{Page Hotness Profiling Mechanism}
There are three popular page hotness profiling mechanisms adopted in prior works (Table \ref{tab:summary_of_literature}):

\textbf{Hint Page Fault Based.}
Ordinary memory accesses are handled by hardware and do not trap into the OS, making them difficult for the OS to observe directly.
Systems therefore use hint-page-fault-based profiling to intercept selected memory accesses and collect access information~\cite{AutoNUMA_lwn, AutoTiering_ATC_21}.
Specifically, this technique periodically scans and makes a range of virtual addresses (VAs) inaccessible by changing their PTE permissions to \texttt{PROT\_NONE} state and flushing their TLB entries.
Consequently, any subsequent access to these pages will trigger a minor page fault (hint page fault).
During the page fault handling, the OS can capture this memory access information, thereby facilitating detailed analysis of memory access patterns.
Note, the process of making pages inaccessible is often called ``unmap'' \cite{AutoNUMA_lwn, unmap_term_pmem, unmap_term_redhat}, but this differs from the ``unmap'' used in page reclamation. In this paper, ``unmap'' specifically refers to modifying PTE permission bits unless stated otherwise.
Also, by periodically scanning the ``unmap'' state, the OS can capture the duration for which a page remains unaccessed, using this as a metric for hotness.

\textbf{PTE Access Bit Based.}
Most modern architectures incorporate an access bit within their PTEs.
The OS can clear this bit, and the Memory Management Unit (MMU) sets it when the corresponding page is accessed.
By periodically scanning and clearing the bit, the OS can infer page access patterns and estimate page hotness.
This enables OS-level access tracking without trapping on every memory access.
For example, Linux uses this mechanism to approximate LRU information for page reclamation~\cite{Nimble_ASPLOS_19}.
However, because the access bit is sampled only periodically and records only whether a page has been accessed since the last scan, it provides coarse-grained hotness information.

\textbf{Hardware Performance Counters Based.}
Modern processors typically include embedded hardware performance counters and support hardware event sampling mechanisms, such as PEBS (Processor Event-Based Sampling) \cite{pebs_event_website} in Intel and IBS (Instruction-Based Sampling)~\cite{IBS_AMD_Research} in AMD processors. Based on this, many works \cite{HeMem_SOSP21, MEMTIS_SOSP_23, TMTS_ASPLOS_23, FlexMem_ATC_24, TPP_ASPLOS23} leverage memory access-related event sampling to collect memory access information for guiding page hotness identification.
However, hardware sampling is probabilistic and often limited by sampling coverage and platform support, which can reduce hotness accuracy~\cite{Chrono_EuroSys2025, NeoMem_MICRO_24}.

\subsubsection{Migration Policy in Tiered Memory System}
The main goal of multi-tiered memory tiering is to efficiently migrate data among tiers based on page hotness. 
In a two-tiered system, migration is straightforward and bidirectional: promotion moves hot pages upward, while demotion moves cold pages downward.
Existing multi-tiered systems such as AutoTiering and MTM extend this style with greedy promotion and stepwise demotion: hot pages are moved to the highest available tier, while cold pages are demoted to a lower-performance tier~\cite{AutoTiering_ATC_21,MTM_EuroSys_24}.
However, this policy can intensify tier contention, making page migration more difficult under capacity pressure.
Related works in two-tiered systems, including FlexMem, PET, and Chrono, address this by proactively demoting cold pages before the top tier experiences memory pressure, keeping space available for future promotions~\cite{FlexMem_ATC_24,PET_EuroSys25,Chrono_EuroSys2025}.
However, directly extending this idea to multi-tiered systems is insufficient: there are multiple possible demotion destinations, and intermediate tiers can still become contention points.

\begin{figure*}[t]
  \centering
  \includegraphics[width=\textwidth]{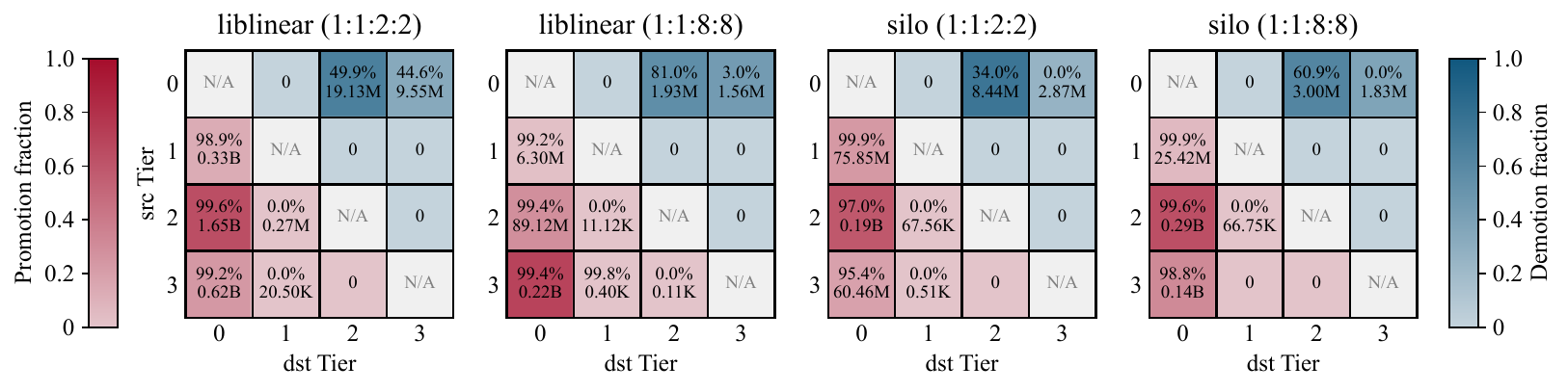}
  \vspace{-0.37in}
  \caption{Migration distribution and failure rates under AutoTiering.
Each cell represents one source (src) and destination (dst) tier pair.
Red and blue cells denote promotion and demotion paths.
Color intensity shows the fraction of total migration attempts on the path, and cell text reports the failure rate and total attempts.
\textit{Note:} AutoTiering excludes demotion from Tier-0 (local DRAM) to Tier-1 (remote DRAM), so the Tier-0$\rightarrow$Tier-1 demotion path has zero observed operations.}
\label{fig:motivation:promotion_and_demotion}
  \vspace{-0.15in}
\end{figure*}

\section{Limitations of Existing Solutions}\label{sec:motivation}
We identify the limitations of existing multi-tiered memory management.
First, limited available tier capacity can hinder promotion and demotion, leading to migration failures.
Second, identical profiling frequency across tiers fail to balance profiling overhead and hotness detection needs.

\subsection{Page Migration Failures in Multi-tiered Systems}\label{sec:motiv:migration_fail}

Effective multi-tiered memory management depends not only on identifying hot pages, but also on whether those pages can be migrated to their destination tiers.
This section quantifies promotion and demotion failures under capacity pressure.
We evaluate AutoTiering~\cite{AutoTiering_ATC_21} as an open-source representative of greedy promotion policy in multi-tiered memory systems.
The evaluation uses Liblinear and Silo under the 1:1:2:2 and 1:1:8:8 tier-capacity configurations, where each ratio denotes the capacity quotas of Tier-0, Tier-1, Tier-2, and Tier-3.
The configured memory capacity is well‑suited to the RSS of the tested workloads.
Detailed configurations are described in \S\ref{sec:eva:woakload_configuration}.
During the runs, we record each promotion and demotion attempt with its source tier and destination tier.
Under greedy promotion, hot pages first attempt to move to Tier-0 and fall back to lower tiers if the target tier is full.
Under stepwise demotion, cold pages are moved downward through the hierarchy, using intermediate tier as possible destination.
As a result, the two migration directions can converge on the same intermediate tier.

Our results indicate that AutoTiering experiences severe promotion failures under both tier-capacity configurations.
Across 1:1:2:2 and 1:1:8:8, promotion failures reach 97.4\% for Silo and 99.4\% for Liblinear.
The path-level results in Figure~\ref{fig:motivation:promotion_and_demotion} show where these failures occur.
We observe that most failed promotions target Tier-0, indicating that the highest tier remains saturated during execution.
Although fallback promotions to lower tiers may succeed, they still leave hot pages in suboptimal tiers and shift migration pressure to intermediate tiers.
Stepwise demotion also uses intermediate tiers as possible destinations for cold pages, which can add to capacity pressure in these tiers.
Across the two configurations, demotion failures reach 25.4\%--37.9\% for Silo and 46.1\%--48.2\% for Liblinear.
For Liblinear, 49.9\%--81.0\% of demotion attempts from Tier-0 (top tier) to Tier-2 (intermediate tier) fail, whereas 99.4\%--99.6\% of promotion attempts from Tier‑2 to Tier‑0 fail, while Silo exhibits a similar result.
Under these policies, intermediate tiers can serve as destinations for both promoted hot pages and demoted cold pages.

Consequently, we identify the first challenge as follows:
\textbf{\textit{Challenge\#1:}}
\textbf{In multi-tiered memory systems, limited available tier capacity can hinder promotions and demotions, leaving pages in suboptimal tiers.}

\begin{figure}[t]
\centering
  \includegraphics[width=0.9\linewidth]{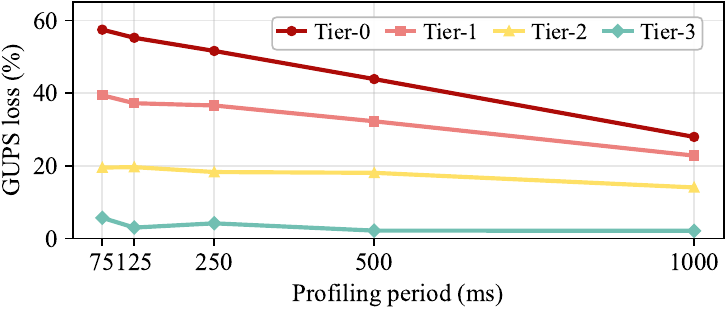}
    \vspace{-0.15in}
    \caption{Profiling period and hotness impact on GUPS loss.}
    \vspace{-0.2in}
\label{figure:motivation:profiling_hotness_period}
\end{figure}

\subsection{Profiling Overhead Varies Across Tiers}\label{sec:motivation:profiling_overhead}

Hardware sampling mechanisms such as PEBS provide useful access information, but their support is often limited and inconsistent across heterogeneous memory tiers~\cite{NeoMem_MICRO_24}.
Access-bit scanning is more portable, but it exposes only a one-bit access state and detects reaccesses only when the page is scanned again.
Hint-page-fault-based profiling is therefore attractive: it is purely software-based, and captures a page reaccess immediately after the page is made inaccessible, providing richer timing memory access information for hotness profiling.
However, the overhead of hint-page-fault-based profiling is not negligible.
First, making pages inaccessible requires costly page table walks (PTWs), PTE permission changing and TLB flushes, which invalidate existing translations.
More importantly, when such pages are accessed again, applications suffer TLB misses, possible PTW, and hint page faults on the critical path.
The performance impact of these costs is mainly exposed when unmapped pages are reaccessed on the application critical path.
Thus, the overhead becomes sensitive to page hotness: hot pages are likely to incur frequent TLB misses and hint page faults, whereas cold pages may remain unaccessed for a long time and impose little performance impact.

To quantify how performance sensitivity to profiling frequency varies across tiers with different access hotness, we run a GUPS-like random update workload~\cite{HeMem_SOSP21} over a 32~GB dataset evenly divided into four 8~GB tiers.
GUPS (Giga-Updates Per Second) is used as the performance metric.
We assign skewed access shares across tiers to set resided pages with different hotness: 64\%, 25\%, 10\%, and 1\% of Tier-0, Tier-1, Tier-2, and Tier-3, respectively.
During each run, we apply profiling to only one target tier by periodically making pages in that tier inaccessible, and decrease the profiling frequency by changing the profiling periods from 75~ms to 1000~ms.
To measure the actual performance overhead, we report GUPS loss, i.e., the percentage reduction in GUPS compared with a baseline run without profiling.

The results are presented in Figure~\ref{figure:motivation:profiling_hotness_period}, where we observe that tiers with higher access hotness are much more sensitive to the profiling frequency.
At 75~ms (highest frequency in this evaluation), profiling Tier-0 reduces GUPS by 57.5\%, while profiling Tier-3 reduces GUPS by only 5.7\%.
The underlying hint-fault counters show the same pattern: at 75~ms, profiling Tier-0 triggers about 32.6 million hint faults, while profiling Tier-3 triggers only 1.78 million.
As the period increases to 1000~ms, Tier-0 loss drops by 29.5\%, from 57.5\% to 28.0\%, whereas Tier-3 changes only from 5.7\% to 2.1\%.
This imbalance leads to the second challenge: \textbf{\textit{Challenge\#2:}}
\textbf{Applying a high profiling frequency can achieve high page hotness detection accuracy, but it introduces significant overhead, especially on upper tiers that contain most of the hot pages, while having less performance impact on lower tiers, since most pages in lower tiers are cold.}

However, existing profiling mechanisms do not provide tier‑specific frequency control, which would be necessary to enable different profiling frequencies on different tiers.
As summarized in Table~\ref{tab:summary_of_literature}, most systems apply an identical profiling rate across tiers.
Some works reduce overhead by profiling only selected tiers, such as lower, remote, or CXL-attached memory~\cite{TPP_ASPLOS23,NeoMem_MICRO_24,AutoNUMA_lwn}.
Such selective profiling is still too coarse-grained for multi-tiered systems because it does not adapt the profiling period separately for each tier.

\section{Design of \myvar~}\label{sec:design}
\subsection{Overview}\label{sec:design:overview}
\begin{figure}[t]
\centering
  \includegraphics[width=0.97\linewidth]{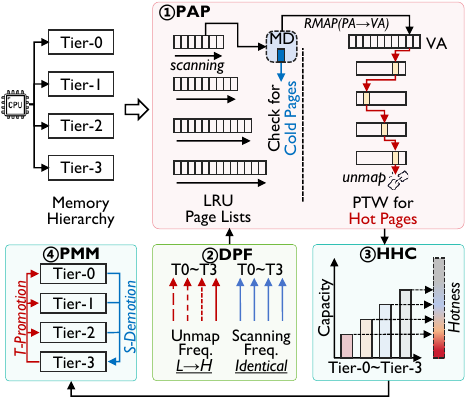}
    \vspace{-0.12in}
    \caption{Overview of \myvar.}
\label{figure:design:overview}
    \vspace{-0.3in}
\end{figure}

Figure~\ref{figure:design:overview} provides an overview of Mole.
Mole aims to provide tier-specific hotness profiling that captures accurate hotness information for all pages with low overhead, and uses this information to drive memory tiering in multi-tiered memory systems.
\myvar~contains four key modules: Physical-Address-based Hotness Profiling (\ding{182} PAP), Differentiated Profiling Frequency Control (\ding{183}DPF), Hotness Histogram Construction (\ding{184} HHC), and Page Migration Management (\ding{185} PMM).
PAP collects page hotness information, while DPF controls the profiling frequency (i.e., the frequency of unmap and scan operations) across tiers.
The collected hotness information is then summarized by HHC into hotness histograms, which guide PMM's page migration.

Four modules are integrated to address the above two challenges.
\textbf{For Challenge\#1}, HHC and PMM reduce tier contention by enabling staging demotion, which bypasses intermediate tiers and moves cold pages directly to the lowest tier, and targeted promotion, which moves hot pages to hotness-matched tiers.
\textbf{For Challenge\#2}, PAP and DPF provide tier-specific profiling with low overhead by selecting pages through physical-page scans and applying different profiling frequencies across tiers.
The rest of this section presents these modules in workflow order: Mole first profiles pages, then summarizes their hotness, and finally performs migration decisions based on the profiling results.

\subsection{Physical Address based Hotness Profiling}\label{sec:design:PAP}
\begin{figure}
    \centering
    \begin{subfigure}[b]{0.23\textwidth}
        \includegraphics[width=\textwidth]{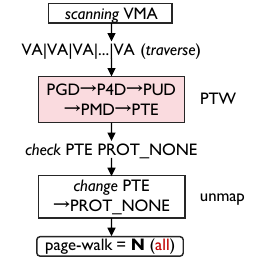}
        \caption{Hint Page Fault}
        \label{fig:design:PAP-1}
    \end{subfigure}
    \begin{subfigure}[b]{0.23\textwidth}
        \includegraphics[width=\textwidth]{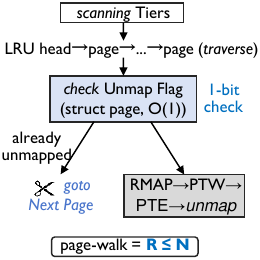}
        \caption{PAP}
        \label{fig:design:PAP-2}
    \end{subfigure}
\vspace{-0.17in}
\caption{Hint page fault based profiling versus PAP.}
\label{fig:design:PAP}
 \vspace{-0.25in}
\end{figure}

\
\indent
\textbf{Profiling limitation.} As discussed in \S\ref{sec:motivation:profiling_overhead}, hint-page-fault-based profiling is attractive because it is software-based and captures page reaccesses promptly after pages are made inaccessible.
However, identifying page hotness in a tier-specific manner still requires periodic scans to check page's location and whether previously unmapped pages remain unaccessed.
However, this process relies on walking VAs and checking page location and PTE states, which introduce costly PTWs, as illustrated in Figure~\ref{fig:design:PAP-1}.

\textbf{Physical page profiling.} To be tier-specific and reduce the overhead, Mole introduces \textbf{P}hysical-\textbf{A}ddress-based Hotness \textbf{P}rofiling (PAP), which scans physical pages directly from Linux's LRU lists and mirror profiling state into page metadata.
This design is motivated by the physical-address-based memory reclamation mechanism of the Linux kernel~\cite{reverse_map_history}.
For each physical page, Mole retrieves the \texttt{struct page} pointer from the LRU list and converts it to the physical address using \texttt{page\_to\_pfn(page) << PAGE\_SHIFT}.
Because Linux maintains separate LRU lists for different tiers, scanning these lists allows Mole to select pages from a specific tier directly.
Mole extends \texttt{struct page}, which serves as metadata for each physical page, with three profiling fields: a 1-bit \texttt{PageUnmap} flag, an unmap timestamp, and a time interval.
The \texttt{PageUnmap} flag records whether the page has been made inaccessible and is synchronized with the PTE permission state whenever Mole performs an unmap operation.
The unmap timestamp records when the page is made inaccessible, and the time interval is used to represent the measured page hotness.
As a result, periodic scans no longer need to walk page tables to check whether the page remains unmapped; they can read the metadata directly.

\textbf{Profiling workflow.} Figure~\ref{fig:design:PAP} compares conventional hint-page-fault-based profiling with PAP.
For a given tier, Mole scans pages in the corresponding LRU list and reads their metadata, as shown in Figure~\ref{fig:design:PAP-2}.
Initially, pages are accessible.
When Mole selects a page for profiling, it performs an RMAP walk (reverse mapping, commonly used in Linux memory reclamation to locate the PTEs mapping a physical page)~\cite{reverse_map_history}, changes the PTE permission to \texttt{PROT\_NONE}, and synchronizes the \texttt{PageUnmap} flag and unmap timestamp in the page metadata.
This unmap operation still requires locating and updating the PTE, but subsequent scans can avoid redundant PTE checks.

After a page is unmapped, Mole handles hot and cold pages differently.
For a hot page, a later access triggers a hint page fault; the fault handler restores the PTE permission, clears the \texttt{PageUnmap} flag, and computes the interval between the unmap timestamp and the fault time as the page's reaccess interval.
For a cold page, if the \texttt{PageUnmap} flag remains set across scan cycles, Mole treats the page as unaccessed and updates its time interval using the current time and the recorded unmap timestamp.
This metadata-only scan records how long the page has remained unaccessed without walking the page table.
In this way, PAP preserves the benefit of hint-page-fault-based profiling for hot pages while reducing scan overhead for cold pages, providing both reaccess intervals and unaccessed durations for later histogram construction and migration decisions.

It is worth noting that VA-based scans can be cache-friendly because adjacent virtual pages map to nearby PTEs and allow direct PTE inspection.
However, many workloads exhibit skewed page hotness, where a large fraction of pages are cold and remain unmapped across multiple scan cycles.
For these pages, repeatedly revisiting their PTEs only confirms that no reaccess has occurred, creating redundant scan overhead.
PAP avoids these repeated PTE checks by reading the \texttt{PageUnmap} flag in page metadata, and further reduces scan cost by processing pages and unmap operations in batches.
As shown later in Figure~\ref{fig:eval:overhead_cpu}, this optimization reduces profiling CPU overhead by 11\% compared with AutoNUMA.

\subsection{Differentiated Profiling Frequency Control}\label{sec:design:DPF}
\
\indent
\textbf{Separating scan and unmap.} As the number of memory tiers grows, finer-grained control over profiling frequency and overhead becomes essential.
Mole controls two profiling operations separately: scan and unmap.
A scan checks page metadata to update cold-page information, while an unmap makes selected pages inaccessible so that later reaccesses can be captured by hint page faults.
Because Linux maintains separate LRU page lists per tier, all pages in a given LRU list belong to the same tier.
Mole leverages this property to apply different profiling controls to pages from different tiers.
The key idea of DPF is to keep low-cost metadata scans frequent across all tiers, while applying expensive unmap operations at tier-specific frequency.

\textbf{Controlling scan frequency.} Mole keeps identical scan frequency across all tiers.
With PAP, a scan normally reads the \texttt{PageUnmap} flag and timing fields in page metadata, without walking page tables or triggering hint page faults.
This low-cost scan continuously updates the unaccessed duration of pages that remain unmapped, helping Mole identify colder pages from every tier.
Therefore, scan frequency can remain high and identical across tiers without imposing the same overhead as frequent unmapping.

\textbf{Controlling unmap frequency and budget.} Mole differentiates the unmap frequency across tiers.
Unmapping may later trigger hint page faults and TLB misses on the application critical path, and this cost is much higher for hot tiers.
Therefore, Mole performs more frequent unmapping on lower tiers, where pages are mostly cold but misplaced hot pages should be detected quickly.
For upper tiers, Mole uses a lower unmap frequency to avoid repeatedly perturbing hot pages.
Let $T$ denote the scan period shared by all tiers.
The unmap period of a tier is set to $m \times T$, where $m \ge 1$ and varies across tiers.
Mole assigns larger $m$ values to upper tiers and smaller $m$ values to lower tiers.

Frequency control determines when each tier performs unmapping, but Mole also needs to bound how much work is performed in each unmap cycle.
Therefore, Mole uses an unmap budget $N$ to limit the total number of pages unmapped in a cycle.
The budget is distributed to each tier in proportion to its capacity.
For example, if Tier-0 accounts for $1/16$ of the total configured memory, Mole unmaps $N \times \frac{1}{16}$ pages from this tier in the corresponding unmap cycle.
As an initialization rule, if a scanned page does not yet have a recorded time interval, Mole forcibly unmaps it once to obtain its first hotness measurement.
This accelerates initial hotness characterization and helps the dynamic thresholds in HHC converge more quickly (see \S\ref{sec:design:HHC}).

\begin{figure}[t]
\centering
  \includegraphics[width=0.92\linewidth]{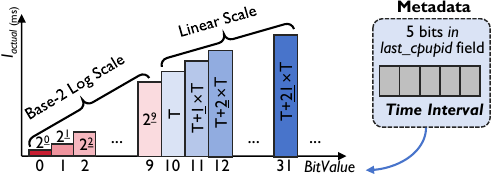}
    \vspace{-0.15in}
    \caption{Time interval representation.}
    \vspace{-0.25in}
\label{figure:design:hotness}
\end{figure}

\subsection{Hotness Histogram Construction}\label{sec:design:HHC}
\
\indent
\textbf{Hotness metric.}
HHC captures both hot-page reaccesses and cold-page inactivity using a unified time-interval metric.
For pages that trigger hint page faults, Mole records the reaccess interval between unmapping and the fault.
For pages that remain unmapped during scans, Mole records the unaccessed duration.
A shorter interval indicates a hotter page, while a longer interval indicates a colder page.

Prior systems such as Chrono also use time intervals for hotness estimation.
Chrono defines Captured Idle Time (CIT) as the interval between page scanning and the subsequent page fault~\cite{Chrono_EuroSys2025}.
Such intervals are effective for hot-page promotion, but they are produced only when an unmapped page is reaccessed.
For cold-page demotion, Chrono still relies on LRU-based selection to choose pages from the fast tier.
Mole extends interval-based hotness to cold pages as well: it records reaccess intervals for pages that fault, and unaccessed durations for pages that remain unmapped across scans.
Compared with LRU-based cold-page selection, this provides a more fine-grained coldness signal and improves the accuracy of selecting pages for staging demotion.

\textbf{Time interval representation.}
Mole stores hotness metadata in \texttt{struct page}, so the representation must be compact to limit per-page memory overhead.
We observe that hot and warm pages often have short, sub-second intervals, while cold pages can remain unaccessed for much longer.
Prior production workload characterization also shows a long cold-page tail in datacenter applications~\cite{TPP_ASPLOS23}.
This wide interval range requires fine granularity for short intervals, while very long intervals only need to be distinguished coarsely once pages are clearly cold.
Although a pure base-2 logarithmic scale is compact, it becomes too coarse for medium and long intervals.
For example, pages with 16--32~s intervals fall into the same bucket, which may mix warm and cold pages and reduce placement accuracy.

To maintain precision with minimal space overhead, Mole adopts the segmented representation shown in Figure~\ref{figure:design:hotness}.
Intervals shorter than \textit{Time\_Threshold} (\textit{T}) use a base-2 logarithmic scale, while longer intervals use a linear scale with step size \textit{T}.
This representation is a space-accuracy tradeoff: it uses limited metadata bits to preserve high resolution for short intervals, while representing longer intervals more coarsely.
\textit{T} is configurable.
Experimentally, we determine that \textit{T}=1024~ms provides a good balance between interval precision and metadata range on our system, and use this setting in our evaluation.
Formally, the recorded interval $I_\text{actual}$ is computed as

\vspace{-0.18in}
\begin{equation}
\text{I}_\text{actual} =
\begin{cases}
\text{BinaryOrder(}\text{I}_\text{ideal}), & \text{I}_\text{ideal} < \text{T} \\
\left\lfloor \log_2 T \right\rfloor + \text{I}_\text{ideal} / T, & \text{I}_\text{ideal} \geq \text{T}
\end{cases}
\end{equation}
\vspace{-0.12in}

where $I_\text{ideal}$ is the real time interval and $I_\text{actual}$ is the recorded interval.
Mole reuses the 21-bit \texttt{last\_cpupid} field from AutoNUMA to store the unmap timestamp and $I_\text{actual}$, allocating 16 bits for the timestamp and 5 bits for the interval.
The 5-bit $I_\text{actual}$ represents values from 0 to 31.
The 16-bit timestamp stores a truncated timestamp to reduce space overhead.

\textbf{Hotness histogram and thresholds.}
Based on the recorded intervals, Mole constructs a hotness histogram and derives capacity-based tier thresholds, following the histogram-based thresholding approach used by MEMTIS~\cite{MEMTIS_SOSP_23}.
Each bucket counts pages in a corresponding interval range, where shorter intervals indicate hotter pages.
Unlike MEMTIS, which uses page access counts, Mole builds the histogram from reaccess intervals and unaccessed durations.

Mole updates the histogram incrementally when a page's recorded interval changes.
Upon a hint page fault, \myvar~immediately synchronizes the page's reaccess interval with the histogram to keep frequently accessed pages up to date.
During scans, \myvar~may also update the histogram using the unaccessed duration of pages that have not faulted.
However, scan-based updates require care because a page that has been unmapped only briefly has not been observed long enough to provide a reliable interval.
Recording such short observations would make the page appear hotter than it actually is.
To avoid this bias, \myvar~updates a page's interval during scans only when the calculated interval exceeds the expected range of its current tier.
Specifically, a daemon periodically derives tier thresholds by walking the histogram from hot to cold.
The daemon accumulates buckets until the page count reaches each tier's capacity boundary, and uses the corresponding intervals as tier thresholds.
These thresholds partition the hotness space according to tier capacities.
Note that Mole slightly reduces the threshold of each tier to avoid capacity constraints, reserving free space to accommodate potential memory underutilization.
PMM then uses the resulting thresholds to decide whether a page should stay in its current tier or migrate to another tier.

\subsection{Page Migration Management}\label{sec:design:PMM}

After PAP and DPF collect page hotness and HHC converts it into tier thresholds, PMM uses these thresholds to guide page migration.
PMM aims to preserve tier capacity for promotions and alleviate the migration difficulties described in \S\ref{sec:motiv:migration_fail}.
Instead of letting promotions and demotions compete for the same intermediate tiers, Mole gives priority to promotion because placing hot pages in faster tiers directly improves performance.
Accordingly, staging demotion moves cold pages directly to the lowest tier, while targeted promotion places hot pages into tiers that match their hotness.

\subsubsection{Staging Demotion}\label{sec:design:PMM:demotion}
\
\newline
\indent
\textbf{Where to demote.}
As illustrated in Figure~\ref{fig:intro:aggressive_demotion}, staging demotion separates demotion destinations from promotion targets.
\myvar~bypasses intermediate tiers and directly demotes pages to the lowest tier.
In this way, the lowest tier serves as a staging area for demoted pages, freeing upper tiers for future promotions.
This design treats demotion primarily as space reclamation, rather than precise placement.
By keeping demoted pages out of upper and intermediate tiers, staging demotion leaves these tiers available for targeted promotion.
Warm pages that are demoted too far can still be detected by high-frequency profiling in the lowest tier and promoted back in time, as described in \S\ref{sec:design:DPF}.

Targeted demotion is a natural alternative that uses HHC thresholds to choose a lower destination tier for each demoted page.
However, it can still occupy intermediate-tier capacity needed by promotions.
Mole therefore prioritizes promotion, which directly improves performance, while treating demotion primarily as space reclamation.
Evaluation shows that targeted-demotion variant (Mole-TD) increases execution time by 2.06$\times$ on average compared with Mole (Figure~\ref{fig:eval:targeted_demotion}).

\textbf{When to demote.}
In conventional \texttt{kswapd}-driven demotion, demotion is tied to Linux's memory reclaim path.
\texttt{kswapd} is woken up only when available pages in a tier fall below the watermark, and it relies on LRU-based reclaim information rather than precise tiering hotness.
As a result, demotion is reactive, and cold pages may be moved only after an upper tier is already under pressure, leaving too little headroom for later promotions.

To avoid this reactive behavior, \myvar~integrates proactive demotion into scan cycles, where it checks unmapped pages and refreshes their unaccessed intervals.
When a refreshed interval indicates that a page is colder than the expected range of its current tier, \myvar~queues the page for asynchronous demotion, leaving headroom for future promotions.
This proactive path handles normal tier maintenance without blocking application execution.
When proactive demotion cannot keep up with urgent memory demand, such as sudden allocations to the preferred upper tier or rapid page-cache growth, \myvar~wakes up \texttt{kswapd} to assist with demotion.
This fallback path also preserves Linux's existing reclaim support for file-backed page-cache pages, whose hotness is not directly tracked by Mole's profiling mechanism.
The destination tier for \texttt{kswapd} demotion is redirected to the lowest tier, so fallback demotion does not occupy intermediate-tier capacity needed by promotions.

\subsubsection{Targeted Promotion}\label{sec:design:PMM:promotion}
\
\newline
\indent
\textbf{Where to promote.}
As illustrated in Figure~\ref{fig:intro:aggressive_demotion}, targeted promotion places hot pages into tiers that match their current hotness.
\myvar~determines the target tier using the dynamic thresholds derived by HHC.
After a page fault, \myvar~computes the reaccess interval between unmapping and the fault, and compares it with the tier thresholds to select a promotion target.
Because intervals are quantized into histogram buckets, a bucket may overlap the boundary between two adjacent tiers.
In this case, \myvar~first promotes the page to the target tier if free space is available; otherwise, it falls back to the next lower tier, where the page can still improve its placement.
If promotion fails, \myvar~defers the current attempt and retries it in the background after asynchronous demotion frees space in the target tier.
\myvar~avoids synchronous demotion because forcing cold-page selection and migration on the fault path would add significant critical-path overhead.

\textbf{When to promote.}
Mole triggers promotions at two points aligned with profiling. First, during the page fault handler, \myvar~identifies hot pages and promotes them if necessary. While this may extend fault handling path, it enables rapid adjustment of data distribution. Second, during scan cycles, \myvar~detects pages whose hotness does not match their current tier. Instead of interrupting the scan to promote each page immediately, \myvar~uses an asynchronous promotion mechanism that batches all identified pages from the current cycle and promotes them together, reducing overhead.

\section{Implementation}\label{sec:implementation}
\
\indent
\textbf{Profiling:} 
We construct a four-tiered memory system, setting the scan period $T$ to 1 second per 1280 MB and the factor $m$ for the lowest tier to 1, based on our evaluation results.
The factor values for other tiers from the sub-lowest to the top are set to 2, 3 and 4. 
Note that \myvar~'s scan frequency is higher than Tiering-0.8, which uses a default scan period of 1 second per 256 MB.
During PA scanning, \myvar~relies on the RMAP walk to enable the unmap operation.
Although each RMAP walk incurs overhead, the frequency of unmap operations can be greatly reduced compared to traditional VA based unmap operations.
This optimization is achieved by flagging the \texttt{struct page} with a \texttt{PageUnmap} bit to filter redundant unmap attempts.
To further reduce overhead, \myvar~employs batched TLB flushing~\cite{batch_flush_tlb} and extends the \texttt{set\_tlb\_ubc\_flush\_pending} mechanism to support deferred invalidation for \texttt{PROT\_NONE} updates~\cite{linux_batched_tlb_inaccessible_pte}.
Notably, \myvar~'s unmap operation is similar to the \texttt{page\_referenced} mechanism in \texttt{kswapd}: both rely on RMAP walk followed by changing PTE permission.
Specifically, \myvar~ scans the \texttt{struct page}, walks the RMAP to locate the PTE, clears the PTE via \texttt{ptep\_get\_and\_clear()}, marks the entry with \texttt{set\_tlb\_ubc\_flush\_pending}, sets the \texttt{PageUnmapped} flag, applies \texttt{PROT\_NONE}, and finally triggers a batched TLB flush. 
Furthermore, the tracking thread runs asynchronously and introduces negligible performance overhead.

\textbf{Metadata and Hotness Histogram:} \myvar~repurposes the existing \texttt{last\_cpupid} field (21-bit) to store unmap timestamp and time interval, maintaining space efficiency.
A possible alternative is to store these fields in \texttt{page\_ext}, which is used by Linux page owner to maintain per-page debugging metadata~\cite{linux_page_owner}.
Mole instead reuses existing \texttt{struct page} fields to keep the profiling state compact and avoid the additional per-page metadata overhead introduced by such extensions.
The \texttt{PageUnmap} flag occupies one bit of the \texttt{struct page} to filter unmap attempts.
To maintain the hotness histogram, \myvar~uses a 32-element counter array per memory tier.
The histogram’s update operations employ optimized synchronization primitives: time interval updates use \texttt{cmpxchg}, while histogram counters rely on per-CPU variables with periodic aggregation into a global atomic counter. Each CPU accumulates local updates and flushes them to the global counter only when a threshold is reached, reducing contention while keeping error within a controllable bound.
The global dynamic thresholds are protected by RCU (Read-Copy-Update) locks to minimize read overhead.
These designs ensure both space and computational overhead remain negligible, with memory usage at just 0.2\%.

\textbf{Page Migration:} Both \myvar's demotion and promotion can be performed asynchronously in the background to minimize performance overhead. 
\myvar's promotion is also triggered synchronously during a page fault, consistent with conventional approaches. 
Although this introduces additional latency in the critical path, it guarantees that hot pages are efficiently migrated to the faster tier.

\textbf{Large Page:} \myvar~is implemented with a default granularity of 4 KB base pages, but its design is agnostic to page size.
First, hint page fault-based profiling remains applicable to large pages.
Second, the hotness identification operates at page granularity, regardless of page size.
Finally, large page migration is naturally supported without requiring additional modifications in Linux.

\section{Evaluation}\label{Evaluation}

We implemented \myvar{} in Linux kernel 5.15, adding kernel support for tier-specific hotness profiling, targeted promotion, and staging demotion.
We evaluate Mole by answering the following questions:

\begin{compactitem}[\scriptsize{$\bullet$}]
    \item How much does Mole improve end-to-end performance and reduce promotion failures? (\S\ref{sec:performance})
    \item How robust is Mole across memory pressure levels, and how do its mechanisms affect migration efficiency and data placement? (\S\ref{sec:performance-reason})
    \item What runtime overhead does Mole introduce? (\S\ref{sec:overhead})
\end{compactitem}

\subsection{Experiment Setup}
We conducted experiments on a dual-socket server platform equipped with Intel(R) Xeon(R) Gold 6330 CPU@2.00GHz processor, where each socket is equipped with 28 cores, 96 GB of DRAM (16 GB DDR4 $\times$ 6) and 256 GB of Intel Optane DCPMM.
These four NUMA memory nodes composed a four-tiered memory system \cite{MTM_EuroSys_24}. 
The memory access latency and bandwidth characteristics of each tier, as measured by Intel's MLC tool \cite{mlc_intel}, are listed in Table~\ref{tab:memory_latency}.
To reduce measurement variability, we disable hardware features such as Hyper-Threading, Turbo Boost, and prefetching.

\begin{table}[t]
    \centering
    \caption{Memory access latency and bandwidth in multi-tiered memory system.}
    \vspace{-0.1in}
    \resizebox{0.45\textwidth}{!}{
        \begin{tabular}{ccc}
        \toprule
        \textbf{Multi-tiered Memory System} & \textbf{Latency} & \textbf{Bandwidth} \\ \midrule
        Local DRAM (1st tier) & 105 ns & 121 GB/s \\ \midrule
        Remote DRAM (2nd tier) & 171 ns & 55 GB/s \\ \midrule
        Local PM (3rd tier) & 327 ns & 5 GB/s \\ \midrule
        Remote PM (4th tier) & 420 ns & 600 MB/s \\ \bottomrule
        \end{tabular}
    }
    \label{tab:memory_latency}
    \vspace{-0.15in}
\end{table}

\begin{table}[t]
\centering
\caption{Evaluated workloads and memory footprints.}
\vspace{-0.1in}
\footnotesize
\setlength{\tabcolsep}{4pt}
\renewcommand{\arraystretch}{0.84}
\begin{tabularx}{\columnwidth}{@{}l c >{\raggedright\arraybackslash}X@{}}
    \toprule
    \textbf{Workload} & \textbf{RSS (GB)} & \textbf{Description} \\
    \midrule[0.08em]
    404.lbm\cite{spec_accel_paper} & 10.1G & Computational fluid dynamics using the lattice Boltzmann method.
    \\
    \midrule
    Graph500\cite{graph500} & 184G & Large-scale graph generation and traversal.
    \\
    \midrule
    Silo\cite{Silo_paper}  & 70.5G & Multicore in-memory database benchmark.
    \\
    \midrule
    Liblinear\cite{liblinear_website}  & 68G & Large-scale linear classification library.
    \\
    \midrule
    GUPS\cite{HPCC_benchmarks, HeMem_SOSP21} & 192G & Random memory access benchmark.
    \\
    \midrule
    PageRank\cite{GAPBS_2017}  & 68G & Graph algorithm measuring node importance.
    \\
    \bottomrule
\end{tabularx}
\vspace{-0.2in}
\label{tab:workload_rss}
\end{table}

\textbf{Workloads.}
Six commonly used workloads, including graph analysis, in-memory databases, HPC, and machine learning, are evaluated in this work~\cite{AutoTiering_ATC_21, HeMem_SOSP21, MEMTIS_SOSP_23}. 
Their RSS (resident set size) details are provided in Table~\ref{tab:workload_rss}.

\textbf{Memory tiering configuration.}\label{sec:eva:woakload_configuration}
Our experimental setup employs tiered memory with capacity ratios of 1:1:2:2 and 1:1:8:8, where the sum of all tier capacities approximately equals the RSS of each benchmark.
For instance, in the 1:1:2:2 configuration, the first two tiers are each set to 1/6 of the RSS, while the last two are each set to 1/3.
To control the capacity of each tier, we configure the capacities of the first two tiers (local DRAM and remote DRAM) using the memmap kernel option~ \cite{memmap_doc}, while managing the latter two tiers (local PM and remote PM) via ndctl~ \cite{ndctl_doc}.
To simulate a vertical four-tiered memory system, we bind the workload exclusively to a single socket (designated as the local socket).

\textbf{Baselines for comparison.} We evaluate \myvar~against three state-of-the-art open-source memory management systems supporting multi-tiered memory: AutoNUMA (also known as NUMA balancing) \cite{NUMA_balancing_lwn}, and Tiering-0.8 \cite{Tiering_git}, AutoTiering \cite{AutoTiering_ATC_21}.
For Tiering-0.8, we enable its AutoNUMA facilities to adapt it to multi-tiered environments.

\begin{figure}[t]
\centering
  \begin{subfigure}{1\linewidth}
    \centering
    \includegraphics[width=\linewidth]{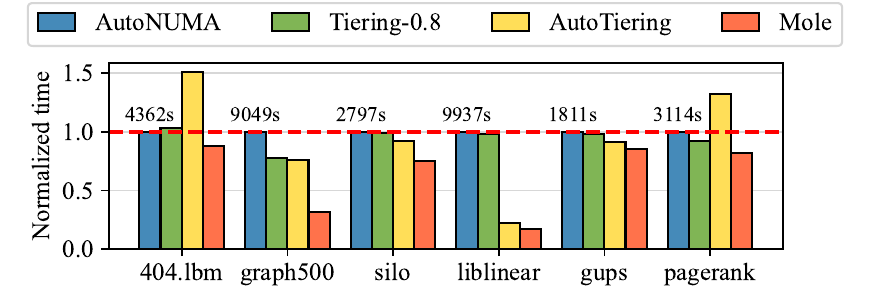}
    \vspace{-0.2in}
    \caption{Tier-0:Tier-1:Tier-2:Tier-3=1:1:2:2}
    \label{fig:eva:result_1_1_2_2}
  \end{subfigure}
  \begin{subfigure}{1\linewidth}
    \centering
    \includegraphics[width=\linewidth]{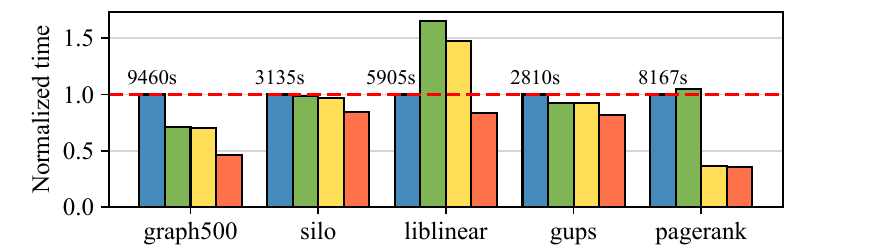}
    \vspace{-0.2in}
    \caption{Tier-0:Tier-1:Tier-2:Tier-3=1:1:8:8}
    \label{fig:eva:result_1_1_8_8}
  \end{subfigure}
  \vspace{-0.3in}
  \caption{Performance under two tiering configurations. 404.lbm is omitted for 1:1:8:8 due to its small RSS.}
  \label{fig:eval:overall_results}
  \vspace{-0.22in}
\end{figure}

\subsection{Performance Comparison}\label{sec:performance}
\begin{figure}[t]
\centering
  \begin{subfigure}{1\linewidth}
    \centering
    \includegraphics[width=\linewidth]{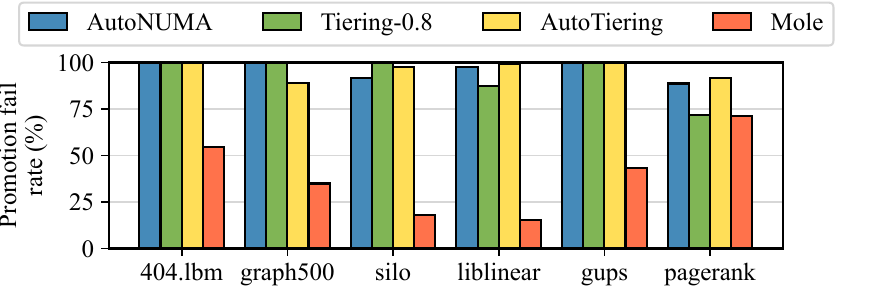}
    \vspace{-0.2in}
    \caption{Tier-0:Tier-1:Tier-2:Tier-3=1:1:2:2}
    \label{fig:eva:promote_1_1_2_2}
  \end{subfigure}
  \begin{subfigure}{1\linewidth}
    \centering
    \includegraphics[width=\linewidth]{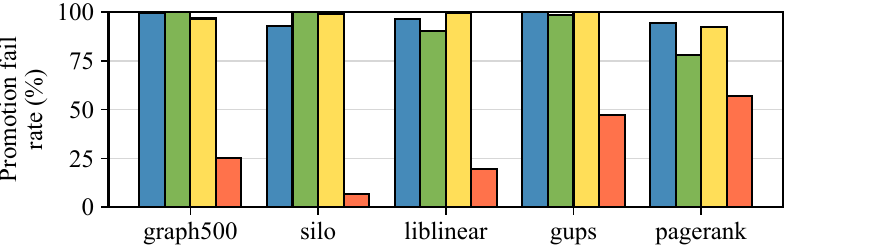}
    \vspace{-0.2in}
    \caption{Tier-0:Tier-1:Tier-2:Tier-3=1:1:8:8}
    \label{fig:eva:promote_1_1_8_8}
  \end{subfigure}
  \vspace{-0.3in}
  \caption{Promotion failure rates under two configurations.}
  \label{fig:eval:promotion_rates}
  \vspace{-0.12in}
\end{figure}

Figure~\ref{fig:eval:overall_results} compares performance in four-tier memory systems under two tiering configurations.
Mole outperforms the second-best system by 70.2\% in the 1:1:2:2 setup and by 20.6\% in 1:1:8:8.
Figure~\ref{fig:eval:promotion_rates} reports promotion failure rates, where Mole reduces the average failure rate to 39.5\% and 31.2\%, while the baselines remain above 90\% on average. This lower failure rate reflects the benefit of separating demotion destinations from promotion targets.

\textbf{SPECaccel.} We use 404.lbm from SPECaccel~\cite{spec_accel_paper}, a lattice-Boltzmann fluid dynamics workload with a small working set.
AutoTiering and Tiering-0.8 fail more than 99.9\% of promotion attempts.
Mole reduces the promotion failure rate to 54.4\% and outperforms the second-best baseline by 14\%.

\begin{figure}[t]
	\centering
    \begin{minipage}{0.45\textwidth}
        \includegraphics[width=\textwidth]{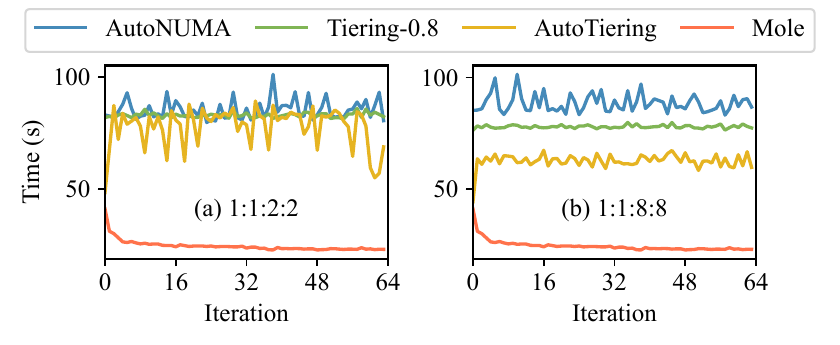} %
    \end{minipage}
    \vspace{-0.2in}
    \caption{Execution time across Graph500 iterations.}
    \label{fig:eval:graph500_iter}
    \vspace{-0.2in}
\end{figure}

\textbf{Graph500.}
We evaluate Graph500 v3.0 with 16 MPI-3 processes~\cite{graph500_github}, using a graph with $2^{28}$ vertices and 6.44 billion edges.
Graph500 alternates between graph construction and traversal, where clustered access patterns make page placement important for later iterations.
Figure~\ref{fig:eval:graph500_iter} shows the per-iteration execution time over 64 iterations.
Under the 1:1:2:2 configuration, AutoTiering and Tiering-0.8 have promotion failure rates of 89.0\% and 99.9\%, respectively, causing consistently higher iteration time than Mole.
Mole reduces the failure rate to 34.9\%, reaches low iteration time after only a few iterations, and outperforms the second-best system by 1.38$\times$.
Under 1:1:8:8, Mole further reduces the failure rate to 25.1\% as the traversal hot set fits better in upper tiers.

\textbf{Silo.}
We evaluate Silo~\cite{Silo_paper} with the YCSB-C workload under a Zipfian distribution, using 400 million key-value pairs and 15 billion lookups (64B keys, 100B values).
The workload exhibits skewed and concentrated access patterns.
Mole reduces promotion failure rates to 17.8\% and 6.8\% under the two tiering configurations, while the baselines remain above 91\%.
This leads to 24\% and 15\% higher performance than the second-best baseline.

\textbf{Liblinear.}
We evaluate Liblinear~\cite{liblinear_website} using the KDD12 dataset~\cite{MEMTIS_SOSP_23}.
Liblinear uses many file-backed pages during initialization and maintains warm page activity during computation.
Under 1:1:2:2, AutoTiering achieves a 4.46$\times$ speedup over AutoNUMA.
However, under 1:1:8:8, Tiering-0.8 and AutoTiering drop to 35\% and 52\% of AutoNUMA, respectively, due to page-cache contention, frequent demotions, and limited cold-page identification accuracy.
AutoTiering also experiences a 48\% demotion failure rate.
Mole's interval-based cold-page tracking supports more effective staging demotion, allowing it to outperform the second-best baseline by 20\% under 1:1:8:8.

\textbf{GUPS.}
We evaluate GUPS with a 192GB memory region, including a 32GB hot subset, performing 1 billion accesses with a 95:5 hot-cold distribution \cite{HeMem_SOSP21}. The hot set roughly matches the top tier, allowing most methods to identify it correctly.
AutoTiering experiences frequent promotion failures when Tier-0 lacks sufficient available capacity, preventing hot pages from reaching the fastest tier.
In contrast, Mole achieves a 56.5\% promotion success rate, yielding a 7.4\% speedup over AutoTiering.

\textbf{PageRank.}
We evaluate PageRank from the GAP suite~\cite{GAPBS_2017} on a Kronecker graph with $2^{28}$ nodes.
PageRank has intensive memory allocation during graph construction and iterative random accesses during traversal.
Although memory pressure becomes moderate during traversal, random accesses still make promotion decisions difficult.
AutoTiering is constrained by static promotion thresholds under 1:1:2:2, while Mole's dynamic thresholds adapt to access patterns.
Mole outperforms the second-best baseline by 12\% and 4\% across the two configurations.

\begin{figure*}[t]
  \begin{minipage}[b]{0.49\textwidth}
    \includegraphics[width=\linewidth]{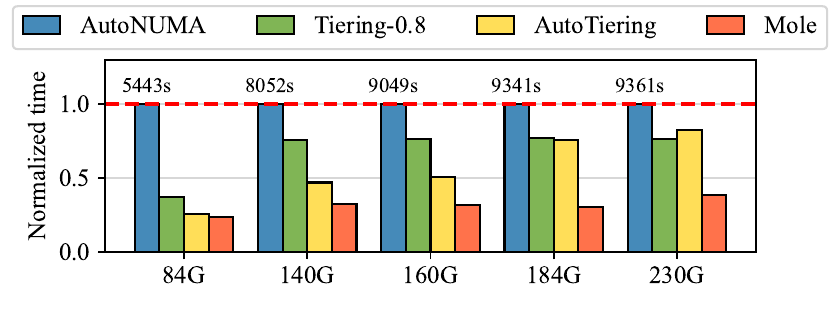}
     \vspace{-0.37in}
    \caption{Performance under fixed tier capacities as workload RSS increases, creating higher memory pressure.}
    \label{fig:eval:large_size_results}
  \end{minipage}
  \hfill
  \begin{minipage}[b]{0.49\textwidth}
    \includegraphics[width=\linewidth]{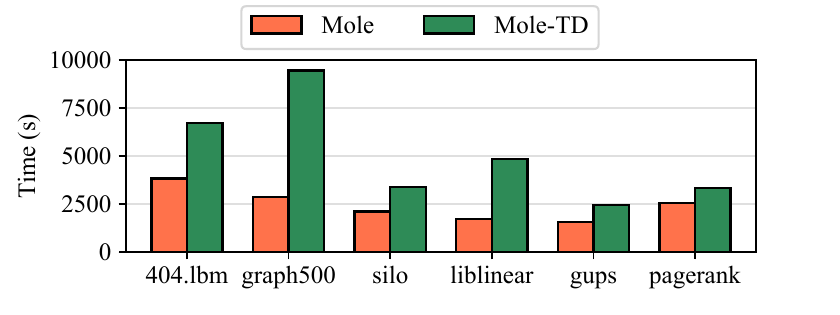}
     \vspace{-0.37in}
    \caption{Performance comparison between Mole and Mole-TD, where TD denotes the variant using \textbf{T}argeted \textbf{D}emotion.}
    \label{fig:eval:targeted_demotion}
  \end{minipage}
  \vspace{-0.1in}
\end{figure*}

\begin{figure*}[h]
  \centering
  \includegraphics[width=0.9\textwidth]{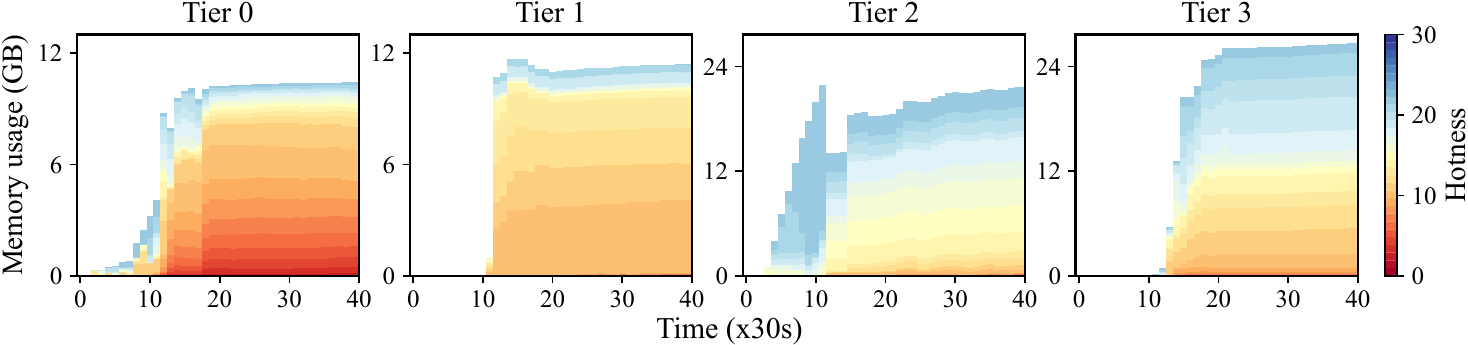}
  \vspace{-0.15in}
  \caption{Memory hotness distribution of Liblinear (1:1:2:2) across tiers over time.}
  \label{fig:eval:liblinear_histogram}
  \vspace{-0.15in}
\end{figure*}

\subsection{Understanding \myvar~Performance}\label{sec:performance-reason}

\subsubsection{Sensitivity to Memory Pressure}\label{sec:memory-pressure}
Under a fixed four-tier memory configuration (Tier-0 : Tier-1 : Tier-2 : Tier-3 = 32GB : 32GB : 64GB : 128GB), we scale Graph500’s RSS from 84 GB to 230 GB. Figure~\ref{fig:eval:large_size_results} shows that Mole achieves on average 2.16$\times$ and 1.74$\times$ speedup over Tiering-0.8 and AutoTiering, respectively.
With small RSS, memory pressure is low and all methods perform similarly.
When the workload’s RSS exceeds the capacity of the first three tiers, Mole’s migration strategy ensures more efficient data placement, sustaining better performance.
Staging demotion is more effective when the lowest tier has available space and holds a small amount of data, as the reduced number of pages to traverse enables Mole to detect hot pages more quickly.

\begin{figure}
	\centering
    \begin{minipage}{0.43\textwidth}
        \includegraphics[width=\textwidth]{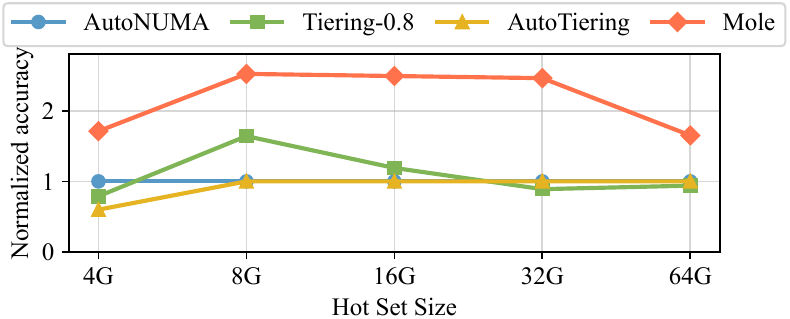} %
    \end{minipage}
    \vspace{-0.125in}
    \caption{Comparison of hot-page identification accuracy.}
    \label{fig:eval:accuracy}
    \vspace{-0.125in}
\end{figure}

\subsubsection{Why Not Targeted Demotion?}
As discussed in \S\ref{sec:design:PMM:demotion}, targeted demotion uses hotness thresholds to choose demotion destinations, but it can consume intermediate-tier capacity needed by promotions.
We implement this as Mole-TD.
As shown in Figure~\ref{fig:eval:targeted_demotion}, Mole-TD increases execution time by 2.06$\times$ on average compared with Mole.
This explains why Mole reserves intermediate tiers for promotion rather than using them as demotion destinations.

\subsubsection{Hotness Distribution Across Tiers.}
\myvar~derives dynamic thresholds from interval-based hotness histograms (see \S\ref{sec:design:HHC}).
Figure~\ref{fig:eval:liblinear_histogram} shows the evolution of Liblinear's hotness distribution under the 1:1:2:2 tiering configuration.
After the initial phase, the hottest pages are concentrated in Tier-0, while staging demotion moves most cold pages to Tier-3.
Although some warm pages remain in Tier-3 under memory pressure, Mole still prioritizes the hottest data for upper tiers, yielding a clear hotness stratification across tiers.

\subsubsection{Hotness Identification Accuracy.}
To evaluate page hotness identification accuracy, we leverage the prior knowledge of the hot-set address range in GUPS and count the number of hot pages identified by each method.
The tier configuration follows the 1:1:2:2 setup, while we vary the hot-set size to examine accuracy under different conditions.
As shown in Figure~\ref{fig:eval:accuracy}, Mole identifies 1.17$\times$, 1.07$\times$, and 1.25$\times$ more hot pages than AutoNUMA, Tiering-0.8, and AutoTiering, respectively.
This improvement mainly comes from Mole's proactive detection of hot pages in lower tiers.

\begin{figure}[t]
    \centering
    \makebox[\columnwidth][l]{%
    \begin{subfigure}[b]{0.22\textwidth}
        \includegraphics[width=\textwidth]{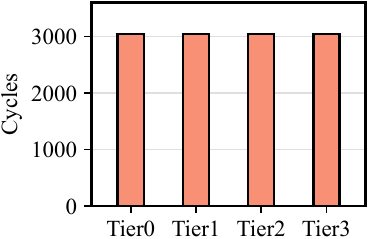}
        \caption{Scan cycles}
        \label{fig:eval:node_unmap_number}
    \end{subfigure}
    \hspace{0.01\textwidth}
    \begin{subfigure}[b]{0.22\textwidth}
        \includegraphics[width=\textwidth]{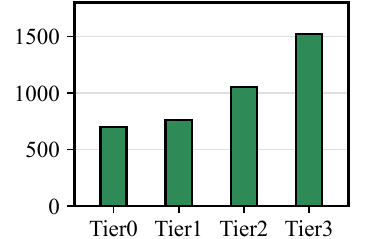}
        \caption{Unmap cycles}
        \label{fig:eval:node_scan_number}
    \end{subfigure}
    }%
  \vspace{-0.15in}
    \caption{Scan and unmap activity across memory tiers.}
    \label{fig:eval:node_scan_and_unmap}
  \vspace{-0.175in}
\end{figure}

\subsubsection{Tier-Specific Profiling Control.}
Figure~\ref{fig:eval:node_scan_and_unmap} shows the cycle distribution across tiers for the GUPS workload with a 1:1:8:8 tier configuration. 
With Mole's differentiated profiling strategy (see \S\ref{sec:design:DPF}), the lowest tier dedicates all its cycles to unmapping, enabling faster hot-page profiling. In contrast, the top tier spends only about 25\% of its cycles on unmapping, primarily relying on scanning to identify warm or cold pages. 
Moreover, over 90\% of pages in the lowest tier are already unmapped due to their cold nature and can be quickly skipped via the \texttt{PageUnmap} flag, reducing the detection latency of hot pages.
For scanning based profiling, as pages in all tiers are asynchronously scanned with an identical frequency, its cycles remain similar across tiers.

\subsection{Overhead Analysis}\label{sec:overhead}
\begin{figure}
	\centering
    \begin{minipage}{0.475\textwidth}
        \includegraphics[width=\textwidth]{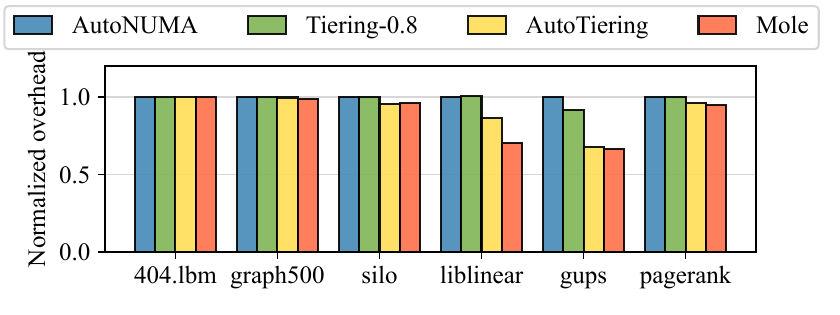} %
    \end{minipage}
    \vspace{-0.2in}
    \caption{Profiling CPU overhead across systems.}
    \label{fig:eval:overhead_cpu}
  \vspace{-0.22in}
\end{figure}

Figure~\ref{fig:eval:overhead_cpu} reports the profiling CPU overhead under the 1:1:2:2 configuration.
Mole reduces CPU overhead by 11\% compared with AutoNUMA, mainly because physical-page scans and the \texttt{PageUnmapped} flag avoid redundant virtual-address-based PTE checks.
Our measurements on Tiering-0.8 show that over 50\% of PTWs revisit PTEs that are already in the unmapped state.
Table~\ref{tab:eval:profiling_overhead} reports the average per-page profiling overhead.
Mole reduces the average operation time from 271.0~ns to 246.7~ns, or 0.91$\times$ that of Tiering-0.8.
Table~\ref{tab:eval:unmap_breakdown} further breaks down Mole's unmap overhead.
The combined PTE update and TLB flush path dominates the remaining cost, accounting for 96.8\%.
In contrast, PTWs and reverse-mapping walks account for only 2.3\% and 0.9\%, respectively.

\begin{table}[h]
\centering
\vspace{-0.05in}
\caption{Per-operation profiling overhead.}
\vspace{-0.1in}
\small
\setlength{\tabcolsep}{8pt}
\begin{tabularx}{\columnwidth}{@{}>{\raggedright\arraybackslash}X
                                >{\centering\arraybackslash}X
                                >{\centering\arraybackslash}X@{}}
\toprule
\textbf{Method} & \textbf{Time (ns)} & \textbf{Normalized} \\
\midrule
Tiering-0.8 & 271.0 & 1.00 \\
Mole & 246.7 & 0.91 \\
\bottomrule
\end{tabularx}
\vspace{-0.15in}
\label{tab:eval:profiling_overhead}
\end{table}

\begin{table}[h]
\centering
\vspace{-0.15in}
\caption{Mole unmap overhead breakdown.}
\vspace{-0.1in}
\small
\setlength{\tabcolsep}{6pt}
\begin{tabularx}{\columnwidth}{@{}
>{\centering\arraybackslash\hsize=0.7\hsize}X
>{\centering\arraybackslash\hsize=0.7\hsize}X
>{\centering\arraybackslash\hsize=1.6\hsize}X
@{}}
\toprule
\textbf{RMAP} & \textbf{PTW} & \textbf{PTE update + TLB flush} \\
\midrule
0.9\% & 2.3\% & 96.8\% \\
\bottomrule
\end{tabularx}
\label{tab:eval:unmap_breakdown}
\end{table}

\section{Related Work}\label{Related Works}

\subsection{Memory Access Tracking Optimization} 
As presented in Table \ref{tab:summary_of_literature}, memory access tracking includes software-based methods~\cite{HotBox_ISMM22, Telescope_ATC_24, MTM_EuroSys_24, HeMem_SOSP21, TMP_IPDPS_21, FlexMem_ATC_24, multi_gen_lwn, damon_lwn, AutoTiering_ATC_21, Thermostat_ASPLOS17, Object_tiering_ISMM2023, Object_tiering_TACO2025}, hardware-enhanced approaches~\cite{NeoMem_MICRO_24, Memstrata_OSDI24, m5_ASPLOS25, Pond_ASPLOS_23, TMTS_ASPLOS_23, DSA_2LM_ATC25, HybridTier_asplos26}, and library-based approaches~\cite{Mira_SOSP23, TeRM_FAST24, XMem_Eurosys16}.
Prior studies have demonstrated that distinct tracking strategies are required for accurately identifying hot and cold pages~\cite{TMTS_ASPLOS_23}.
TPP~\cite{TPP_ASPLOS23} restricts tracking to only a single tier in two-tiered memory systems, which is too coarse-grained and inadequate for the complexity of multi-tiered memory architectures.
AutoTiering~\cite{AutoTiering_ATC_21}, although tracking across all tiers, is configured to focus on cold page detection by default, maintaining hotness information only for the upper tiers.
The strategy may result in limited responsiveness to page hotness changes in lower tiers.
TMTS~\cite{TMTS_ASPLOS_23} adopts a tier-aware strategy, but its tier-awareness is limited: its periodic access-bit scanning is tier-agnostic, as it scans VAs across all tiers. Its access sampling on the capacity tier relies on restricted PEBS events to filter top-tier traffic, but these events capture only loads, lack store support, and are unavailable on many tiers and platforms, which limits scalability in multi-tiered systems. In contrast, Mole uses a purely software-based profiling approach that operates uniformly across all tiers and platforms.

\subsection{Page Hotness Identification}
Prior works have enhanced page hotness identification~\cite{AutoTiering_ATC_21, Colloid_SOSP24, TieredMMS_MSST23, refault_latency_BigComp22, TMTS_ASPLOS_23, MULTI_CLOCK_HPCA22, Chrono_EuroSys2025, IDT_HPDC_24}, and some focused on virtualization and cloud-oriented scenarios~\cite{TMO_ASPLOS22, vTMM_EuroSys23, Memstrata_OSDI24, Demeter_SOSP25}.
MEMTIS~\cite{MEMTIS_SOSP_23} implements an access histogram based dynamic threshold, but its effectiveness is constrained by the randomness of PEBS sampling~\cite{Nomad_OSDI24}.
Besides, MEMTIS relies on access frequency as its metric, requiring multiple samples to determine a page’s hotness and periodic cooling of all pages, which limits its sampling frequency.
Tiering-0.8~\cite{Tiering_git} and Chrono~\cite{Chrono_EuroSys2025} apply time-interval-based tracking only for hot page, falling back to simple LRU for demotion, limiting their effectiveness in multi-tier settings. Mole evaluates a page’s unaccessed duration prior to hint page fault, providing the accurate cold-page identification needed for efficient demotion.

\subsection{Page Migration}
There are also some works focused on optimizing page migration mechanisms~\cite{Nomad_OSDI24, HiNUMA_ICCD_2019, Page_migration_support_MEMSYS19, Memos_TPDS19, Nimble_ASPLOS_19, Colloid_SOSP24, AMP_TC_2022, UniMem_ATC24}.
AutoTiering~\cite{AutoTiering_ATC_21} emphasizes the tier-aware memory model and introduces a new migration path for multi-tiered memory systems. However, its greedy promotion migration strategy leads to suboptimal tier utilization and performance degradation under memory pressure.
FlexMem~\cite{FlexMem_ATC_24} increases the demotion rate to free space for promotions, while PET~\cite{PET_EuroSys25} and Chrono~\cite{Chrono_EuroSys2025} employ proactive demotion to improve performance by acting before memory pressure occurs. These approaches primarily optimize demotion rate and trigger timing, which is sufficient for two-tiered systems but does not address the tier contention problem in multi-tiered systems.

\section{Conclusion}\label{sec:Conclusion}
In this work, we present Mole, a memory management system for multi-tiered memory hierarchies.
Mole alleviates tier contention with staging demotion, which moves cold pages directly to the lowest tier, and targeted promotion, which places hot pages into tiers that match their hotness.
These mechanisms are enabled by tier-specific hotness profiling, allowing Mole to track page hotness with low overhead.
This enables timely detection of reactivated hot pages while limiting the profiling overhead imposed on upper tiers.
Evaluation on a four-tier memory system shows that Mole outperforms SOTA tiering systems.

\normalem
\bibliographystyle{plain}
\bibliography{ref.bib}

\end{document}